\documentclass[runningheads]{llncs}
\usepackage[T1]{fontenc}
\usepackage{graphicx}

\usepackage{subcaption}
\usepackage{tabularx}

\begin{document}

\title{Beyond Satisfaction: From Placebic to Actionable Explanations For Enhanced Understandability}
\titlerunning{Beyond Satisfaction}

\author{Joe Shymanski\orcidID{0009-0007-7955-192X} \and Jacob Brue\orcidID{0009-0001-6656-3960} \and Sandip Sen\orcidID{0000-0001-6107-4095}}
\authorrunning{J. Shymanski \textit{et al.}}

\institute{The University of Tulsa, Tulsa OK 74104, USA
\email{\{joe-shymanski,jacob-brue,sandip-sen\}@utulsa.edu}
}

\maketitle

\begin{abstract}
Explainable AI (XAI) presents useful tools to facilitate transparency and trustworthiness in machine learning systems. However, current evaluations of system explainability often rely heavily on subjective user surveys, which may not adequately capture the effectiveness of explanations. This paper critiques the overreliance on user satisfaction metrics and explores whether these can differentiate between meaningful (actionable) and vacuous (placebic) explanations. In experiments involving optimal Social Security filing age selection tasks, participants used one of three protocols: no explanations, placebic explanations, and actionable explanations. Participants who received actionable explanations significantly outperformed the other groups in objective measures of their mental model, but users rated placebic and actionable explanations as equally satisfying. This suggests that subjective surveys alone fail to capture whether explanations truly support users in building useful domain understanding. We propose that future evaluations of agent explanation capabilities should integrate objective task performance metrics alongside subjective assessments to more accurately measure explanation quality. The code for this study can be found at https://github.com/Shymkis/social-security-explainer.
\end{abstract}

\section{Introduction}

People provide explanations while communicating for a wide range of purposes. Explanations can communicate understanding, experience, perspective, and much more. People expect similar types of communication from AI systems they interact with. Since such digital systems function in fundamentally different ways than humans, explanations are often needed to increase the understandability or interpretability of the behavior of the digital system itself. Hence, there has been an increased focus on providing explainability for AI systems in human-AI interactions. With continuous progress towards larger, deeper, and more complex AI models, these systems can reach significant complexity, increasing the difficulty of providing transparency and interpretability of actions taken or recommendations provided by these systems. The field of XAI faces the challenge of generating high-quality explanations to increase the understandability of AI models.

To guide the creation of better explanations and to measure their effectiveness, we should select appropriate evaluation metrics to evaluate the quality of the explanations. There is an extensive body of research within the social sciences that studies human explanations~\cite{miller2019explanation}. Explanations can be used to transfer the knowledge and perspective of the explainer. Effective explanations follow the rules of conversation by being both minimal and useful. They also take into account the explainer's biases and expectations about the explainee. 

There has been much work on designing computer-centered evaluation metrics based on the principles of quality explanations found through the social sciences. These metrics do not involve user studies, but can be calculated based on the explanations themselves~\cite{mohseni2021multidisciplinary}. Most of these metrics are built around measuring model fidelity, interpretability, or simplicity~\cite{lopes2022xai,schwalbe2023comprehensive}. Although these theoretical principles provide useful guidelines, since a real user is never involved in the evaluation process, they do not reliably indicate the effectiveness of the explanation on their own and can be misleading~\cite{chowdhury2024objective,keane2021if}.

Many studies rely on user satisfaction reported to assess explanation effectiveness. Others choose to test their users' mental models through their ability to act on the understanding provided by the explanations~\cite{chromik2020taxonomy,mohseni2021multidisciplinary,schwalbe2023comprehensive}. The methods for measuring the effectiveness of the explanation vary widely from one study to the next. There is no agreed upon standard for how explanations should be evaluated, so choosing the correct criteria for the goal of the XAI system is paramount. We subsequently aim to demonstrate the pitfalls of a common evaluative disconnect in the literature---overreliance on subjective satisfaction metrics over objective user actions and outcomes.

To do this, we introduce the notion of \textbf{placebic explanations}—--responses that appear explanatory but fail to improve understanding. Despite containing surface-level information, such explanations often restate known facts or tautologies without revealing model reasoning. Their deceptive effectiveness in past social science experiments raises concern for modern AI systems~\cite{langer1978mindlessness}. On the other hand, \textbf{actionable explanations} carry meaning that has the potential to increase the understandability of the agent. We tested user satisfaction and user performance in the presence of placebic and actionable explanations.

We designed a Social Security filing age optimization study to compare evaluation methods for explanations of varying qualities. The study allows users to learn about the optimal filing age of Social Security through explanations received along with the correct decision for various representative scenarios presented. The study includes user groups that received no explanations, placebic explanations, and actionable explanations.

\section{Related Works}

\subsection{Explanations in the Social Sciences}

Although automated explanation generation systems are a relatively modern technology, there is a mature background of study in the social sciences on human explanations. 

In his pivotal work, "Explanation in artificial intelligence: Insights from the Social Sciences", Miller draws several conclusions about the goals of XAI from research within the Social Sciences~\cite{miller2019explanation}. He finds that explanations, among other things, should be conversational, contrastive, and useful. We use these guidelines to produce actionable explanations for our study.

Langer {\em et al.} introduced the concept of placebic information when they showed that people were willing to accept less informative excuses for smaller inconveniences. They proposed that people would only consider the content of the explanation with care if there was a triggering reason for the consideration. They found that people would notice when an explanation is not given, but may be satisfied by a placebic explanation~\cite{langer1978mindlessness}.

\subsection{Meta-analysis in XAI}

Van der Waa {\em et al.} identify a lack of focus on evaluation within articles covering studies of XAI methods~\cite{van2021evaluating}. Many of them did not evaluate with their users and many relied heavily on subjective measures of satisfaction, trust, and confidence. This severely limits the ability of researchers to compare XAI methods between studies.

In their review of XAI methods and taxonomy, Lopes {\em et al.}~\cite{lopes2022xai} found that "a large portion of the literature follows a qualitative evaluation of user satisfaction of explanations, using questionnaires and interviews." This measure is very direct, since the goal of explanations is to provide satisfaction to users. However, they note that this is more useful for experts who have more knowledge about the subject and the system than for lay people whose satisfaction may not be representative of explanation quality. Model understandability through testing the user's mental model is another metric used by authors, and this method is less subjective and more comparable across studies than subjective user reported metrics.

DARPA developed an XAI program to promote explainability for AI systems~\cite{gunning2019darpa}. They identify a variety of measures for explanation effectiveness, including mental model understanding and explanation satisfaction. Their directions and guidelines have been an important focal point for the field of XAI.

Hoffman {\em et al.} identify the essential need to determine content validity to validate the scale provided for the satisfaction rating~\cite{hoffman2018metrics}. They performed a Content Validity Ratio (CVR) for nine questions for use in determining user satisfaction. We use a similar list of questions with a narrower focus on the feeling of satisfaction.

Following the explosive growth of the XAI field, several survey papers have covered the various approaches taken across the field. In Gesina Schwalbe and Bettina Finzel's survey of XAI surveys, separate explainable AI metrics into three categories~\cite{schwalbe2023comprehensive}. Functionally-grounded metrics, like faithfulness and stability, do not require user involvement to produce. Human-grounded metrics, such as the accuracy and efficiency of mental models, perform measurements from human users. Application-grounded metrics require the user to experience the explanation within the context of the final application.

Mohseni {\em et al.} surveyed the field of XAI, paying close attention to the different design goals of XAI and the methods for evaluating XAI~\cite{mohseni2021multidisciplinary}. Two prominent targets include the user's mental model and the explanation's usefulness and satisfaction to the user. They found that usefulness and satisfaction were most commonly utilized for the goals of producing algorithm transparency, user trust and reliance, and model debugging. We are most concerned with goals that help nonexperts increase their understanding.

Chromik and Schuessler created a taxonomy for the evaluation of Black-Box explanations in XAI~\cite{chromik2020taxonomy}. They identified three "levels" for evaluation: satisfaction, comprehension, and team performance. We focus on satisfaction and comprehension goals, since team performance is not a goal for these explanations.

Buçinca {\em et al.} argue that human AI teams should not use subjective measures or mental model performance~\cite{buccinca2020proxy}. Since neither is a good predictor of the effectiveness of explanation in human-agent teams, they recommend the team performance metric. This highlights the importance of fitting the metrics to the goals of the system. We focus on explanations where an increased understanding of the system's operation is the primary purpose for the explanations.

Chromik {\em et al.} found that users receiving explanations, especially non-domain expert users, experienced an illusion of understanding when presented with insufficient local explanations~\cite{chromik2021think}. This bias was reflected in subjective user ratings of their own understanding. This rating decreased after the users participated in a mental model test.

\subsection{Explanation Types}

\subsubsection{Actionable Explanations}

Poyiadzi \textit{et al.} argue that explanations should focus on actionable counterfactual explanations that focus on features that are changeable by a user~\cite{poyiadzi2020face}. Similarly, Rasouli and Yu in their article emphasize the value of personalized explanations based on actionable recourse, focusing on elements within the control or priority of the user~\cite{rasouli2024care}. 

We use actionable explanations to describe a broader class of explanations---explanations that actively introduce new information dependent on the underlying model which may lead to increased understanding. 

\subsubsection{Placebic Explanations}

Within the field of XAI, placebic explanations are relatively underutilized. Very few studies use this as a comparison for their explanations. The studies that chose to use placebic explanations made significant conclusions during their analysis compared to actionable explanations. 

Eiband \textit{et al.} studied user's reported trust in a healthy choice recommendation system. Users perceived the system as more trustworthy when a "real" explanation was provided with the recommendation. However, they also found the same change in trustworthiness for "placebic" explanations~\cite{eiband2019impact}.

Nourani \textit{et al.} performed a study for a vision classification task. They provided a visual explanation in the form of a heatmap. They demonstrated a significant difference between the confidence of users who received meaningful explanations as compared to those who received meaningless explanations~\cite{nourani2019effects}.

Bingjie Liu studied the user's perception of decisions based on a variety of factors, including perceived human involvement. They compared groups that received no explanation to placebic explanations and "transparent" explanations. They found that placebic explanations reduced uncertainty and increased trust when the explanation was perceived to come from a human, but not a machine~\cite{liu2021ai}.

Lee \textit{et al.} performed a user study on the perception of bias within moderation and removal decisions. The study finds that placebic explanations decrease the perception of bias for human moderators, but not for AI moderators~\cite{lee2022something}.

Pias \textit{et al.} explored how user groups with different personality features experience explanations differently. They found that placebic explanations strongly improved agreement with the AI classifier over absence of explanation~\cite{pias2024drawback}.

Ehsan and Riedl examined the potential harmful pitfalls that can arise within XAI solutions. They found that placebic explanations could cause misplaced trust in users~\cite{ehsan2024explainability}.

This work follows our previous research in comparing placebic and actionable explanations~\cite{shymanski2025not}. In that study, we enlisted Amazon MTurk workers to complete chess puzzles, some of whom were given explanations from an AI assistant to help them learn the task. We found that in the chess puzzle domain, users performed significantly better on tests when receiving actionable explanations than when receiving placebic explanations. In this work, we test the effectiveness of satisfaction ratings and testing performance in a domain that is simpler, with a lower variance in prior knowledge among experimenters.

\section{Hypotheses}

We believe that actionable explanations inherently boost model understandability, resulting in better user performance. However, placebic explanations lack this quality by definition, so they should result in user performances that are no better than those without any explanations at all.

At the same time, we believe that both types of explanations will similarly satisfy users, as they may perceive the mere presence of an explanation as the system’s attempt to clarify the complexity of the model. However, when explanations are not provided, we anticipate greater confusion, frustration, and lower satisfaction ratings.

For these reasons, we have developed the following hypotheses:
\begin{description}
    \item[H1:] Actionable explanations lead to significantly better user performance compared to placebic explanations.
    \item[H2:] Placebic explanations do not lead to significantly better user performance compared to the absence of explanations.
    \item[H3:] Actionable explanations do not lead to significantly higher user satisfaction compared to placebic explanations.
    \item[H4:] Placebic explanations lead to significantly higher user satisfaction compared to no explanations.
    \item[H5:] Actionable explanations do not lead to significantly higher perceived explanatory power compared to placebic explanations.
\end{description}

\section{Methodology}

\subsection{Domain}

\begin{figure}
    \centering
    \includegraphics[width=\textwidth]{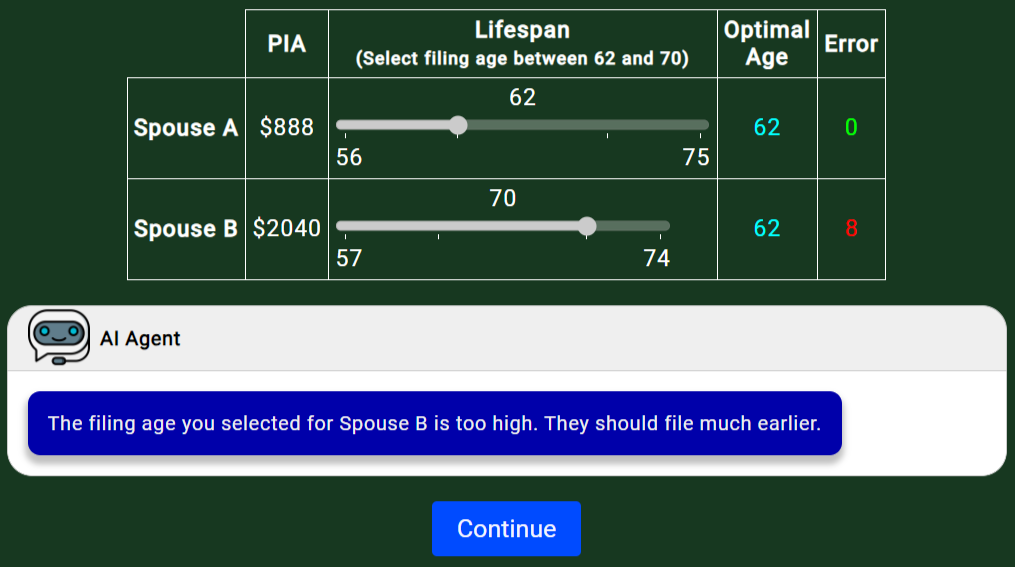}
    \caption{User interface in practice section with a placebic explanation.}
    \label{fig:screenshot_practice}
\end{figure}

We presented participants with a series of hypothetical Social Security scenarios (Figure~\ref{fig:screenshot_practice}), with the goal of optimizing the total payout in each scenario by selecting the best filing ages for the individuals involved~\cite{opensocialsecurity}. Each scenario incorporated several variables for participants to consider: marital status, primary insurance amount (PIA), current age, and lifespan.

The ground truth model comes from opensocialsecurity.com, a calculator application that uses factors like lifespan, marital status, and income to produce inflation adjusted expected value for different filing decisions~\cite{opensocialsecurity}. Actionable explanations were created by researchers, and were designed to accurately reflect the underlying model. In contrast, placebic explanations were designed to include no new information. Example explanations are available in Appendix~\ref{sec:example_explanations}

We chose this domain for several reasons. First, the task was inherently complex and challenging to learn without assistance, while simple explanations could facilitate better decision-making. Second, user performance could be objectively measured using an error metric that compared selected filing ages with the optimal ones. Third, the task was unfamiliar to the vast majority of participants, ensuring that most entered the study with a similar low level of expertise. Lastly, the real-world relevance of social security optimization added an incentive for participants to engage in the task and learn from the information provided.

\subsection{Study Design}

This study followed a between-subjects design with three protocols: no explanations (\textit{None}), placebic explanations (\textit{Placebic}), and actionable explanations (\textit{Actionable}).

We enlisted participation through Prolific, an online platform for recruiting vetted research subjects. Although there are many alternatives, such as Amazon Mechanical Turk or CloudResearch, Prolific appeared to offer the most engaged participants~\cite{albert2023comparing,douglas2023data,eyal2021data,palan2018prolific}. We filter our participant pool to only those who resided in the United States, had at least a 98\% approval rate, and were at least 18 years old.

Participants were sent to our study's website and entered their Prolific ID, which we used to ensure there were no duplicates. Then they chose whether to consent to the study conditions. Those who answered affirmatively moved on and subsequently completed a short demographic survey. Attention checks were used throughout our surveys to ensure all participants were sufficiently focused on the study.

After these initial steps, users entered the practice section, where they were asked to complete 10 social security filing decision tasks in 7.5 minutes. If the participant belonged to the \textit{None} group, they received no help from the AI agent. The other two groups received explanations from the agent after each scenario, enabling participants to read feedback before the next task. Appendix~\ref{sec:example_explanations} provides some examples for each explanation type in different scenarios.

Once they finished the practice section, our subjects moved on to the testing section, where they were presented with 10 more optimization scenarios and the same time limit. Here, none of the groups received help from the agent, and all groups were rewarded with bonus compensation for their accuracy in each selection.

After the testing section, our subjects entered the final survey, where they rated their agreement with several statements on a 7-point Likert scale. We evaluated three subjective metrics with these statements: satisfaction with the practice section, satisfaction with the agent, and the explanatory power of the agent. Each metric was calculated by averaging the responses of a participant to three similar statements. The exact statements in this survey can be found in Appendix~\ref{sec:survey_statements}.

\subsection{Interface}

Figure~\ref{fig:screenshot_practice} shows an example of the user interface for the study practice section. Each scenario was given as a table where one row represented one individual and each column represented the relevant variables. Initially, the PIA and lifespan columns would be populated with data, while the optimal-age and error columns would be left empty. Problems with only one row signified an unmarried situation, whereas those with two rows constituted a married scenario (hence the inclusion of the word "Spouse").

In addition to what is shown in Figure~\ref{fig:screenshot_practice}, a box of definitions was provided in the sidebar for all users. It contained short descriptions of relevant terms, such as PIA and spousal benefits. In addition, the top of the page contained the name of the section, a brief set of instructions, a countdown timer, and a progress bar.

The users would examine the values given to them and move the sliders to the desired filing ages for each individual. Each slider represented a single selection in the scenario. To lock in their answers, users would click the blue submit button beneath the current problem. The optimal ages and errors would then appear. If the subject was in either the \textit{Actionable} or \textit{Placebic} group, the AI agent's text box would populate with at least one explanatory message. These explanations differed between scenarios and protocols. Meanwhile, the \textit{None} group would not see such a text box. After a short delay, the user could click the continue button to move on.

The testing section looked similar to the practice section, though with a few key differences. No explanations were provided to any of the participants and the box of definitions from the practice section was removed. In addition, the bonus compensation earned for each selection was provided in a new column next to the error; it was also filled upon submission.

\section{Results}

We recruited 189 participants from Prolific and divided them equally into three groups of 63, corresponding to the three protocols. In particular, this sample size is significantly larger than that of previous similar studies~\cite{eiband2019impact,langer1978mindlessness}.

\subsection{Demographics}

\begin{figure}
    \centering
    \begin{subfigure}{.49\textwidth}
        \centering
        \includegraphics[width=\textwidth]{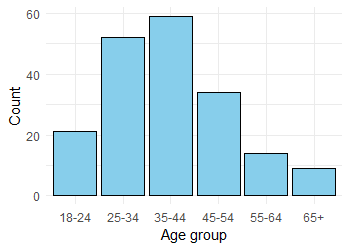}
        \caption{Age.}
        \label{fig:dist_age}
    \end{subfigure}
    \begin{subfigure}{.49\textwidth}
        \centering
        \includegraphics[width=\textwidth]{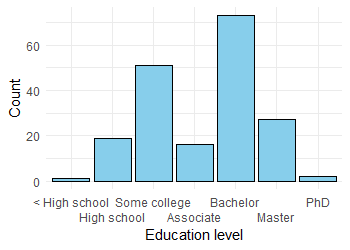}
        \caption{Education.}
        \label{fig:dist_education}
    \end{subfigure}
    \begin{subfigure}{.49\textwidth}
        \centering
        \includegraphics[height=.45\textwidth]{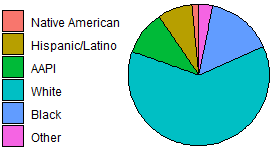}
        \caption{Ethnicity.}
        \label{fig:dist_ethnicity}
    \end{subfigure}
    \begin{subfigure}{.49\textwidth}
        \centering
        \includegraphics[height=.45\textwidth]{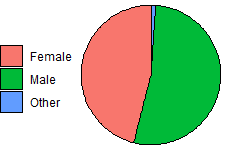}
        \caption{Gender.}
        \label{fig:dist_gender}
    \end{subfigure}
    \begin{subfigure}{.49\textwidth}
        \centering
        \includegraphics[width=\textwidth]{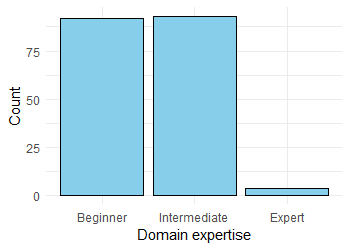}
        \caption{Expertise.}
        \label{fig:dist_skill}
    \end{subfigure}
    \caption{Demographic distributions.}
    \label{fig:dist_demographics}
\end{figure}

Before completing any Social Security tasks, participants filled out a short demographics survey. In it, we requested each participant's age group, education level, ethnicity, and gender, all of which are displayed in Figure~\ref{fig:dist_demographics}. The age distribution of our participants can be found in Figure~\ref{fig:dist_age}. Most of the subjects were between the ages of 25 and 44. Additionally, Figure~\ref{fig:dist_education} shows the various education levels obtained by our users. Most of the respondents experienced some college or graduated with a Bachelor's degree. The distributions of ethnicity and gender are modeled by Figures~\ref{fig:dist_ethnicity} and~\ref{fig:dist_gender}, respectively. Overall, Prolific was able to provide us with a representative sample of US-based subjects.

All subjects indicated their knowledge of the domain, displayed in Figure~\ref{fig:dist_skill}. The minuscule number of self-described experts supports the claim that all participants were similarly inexperienced with Social Security payout optimization.

\subsection{Performance}

\begin{figure}
    \centering
    \begin{subfigure}{.49\textwidth}
        \centering
        \includegraphics[width=\textwidth]{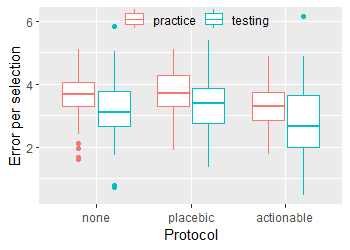}
        \caption{Error.}
        \label{fig:boxplot_error}
    \end{subfigure}
    \begin{subfigure}{.49\textwidth}
        \centering
        \includegraphics[width=\textwidth]{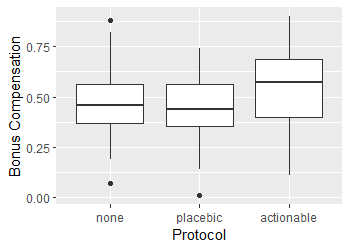}
        \caption{Bonus compensation.}
        \label{fig:boxplot_bonus}
    \end{subfigure}
    \caption{Distributions of both objective performance metrics across protocols.}
    \label{fig:boxplot_performance}
\end{figure}

We utilized two objective metrics to evaluate user performance: test error and bonus compensation. The users faced 10 scenarios in the testing section with a total of 18 selections, all of which were evenly weighted. The bonus for a single selection decreased by 25\% the maximum value for each year of error. Thus, an error of four or more years would yield no bonus compensation for that selection. Figure~\ref{fig:boxplot_performance} shows the distributions of error and bonus compensation for each protocol.

\begin{figure}
    \centering
    \begin{subfigure}{.49\textwidth}
        \centering
        \includegraphics[width=\textwidth]{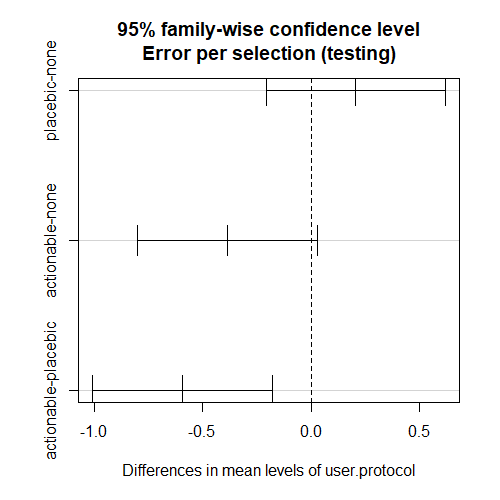}
        \caption{Testing error.}
        \label{fig:tukey_error}
    \end{subfigure}
    \begin{subfigure}{.49\textwidth}
        \centering
        \includegraphics[width=\textwidth]{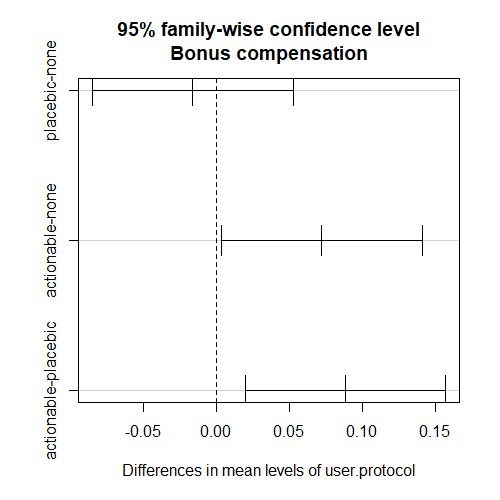}
        \caption{Bonus compensation.}
        \label{fig:tukey_bonus}
    \end{subfigure}
    \caption{Tukey's HSD results for both objective performance metrics.}
    \label{fig:tukey_performance}
\end{figure}

We performed a one-way analysis of variance (ANOVA) to determine if there was a difference in the average testing errors between the three protocols. It revealed the existence of a significant difference among the groups ($F(2, 186)=5.893$, $p=.0033$, $\eta^2=.06$). We then ran a post-hoc analysis using Tukey's honestly significant difference (HSD) test. It found that the mean testing error of the \textit{Actionable} group was significantly smaller than that of the \textit{Placebic} group ($p=.0025$). The \textit{Actionable} group did not have a significantly smaller mean than the \textit{None} group ($p=.0733$), and there was no significant difference between \textit{Placebic} and \textit{None} ($p=.4672$). Figure~\ref{fig:tukey_error} displays these Tukey's HSD results.

We then assessed the difference in mean bonus compensation between the protocols using ANOVA. This also revealed a significant difference ($F(2, 186)=5.255$, $p=.0060$, $\eta^2=.05$). Tukey's HSD discovered that the \textit{Actionable} group received significantly higher bonuses than the \textit{Placebic} ($p=.0075$) and \textit{None} ($p=.0362$) groups. Again, it found no significant difference between \textit{Placebic} and \textit{None} ($p=.8452$). These results are shown in Figure~\ref{fig:tukey_bonus}.

We also ran two one-sided $t$-tests (TOST) on the performances of each pair of protocols to see if any were significantly the same. For testing error, we set the equivalence bound (EqB) to 0.5, since errors could range from 0 to 8. For bonus compensation, the range of possible values was much smaller (0 to 1), so we set its EqB to 0.1. Across both metrics, we found that the performances of the \textit{None} and \textit{Placebic} users were significantly equivalent ($t(124)=1.821$, $p=.036$, Hedges' $g=-.23$ and $t(124)=-3.137$, $p=.001$, Hedges' $g=.11$).

\subsection{Learning}

We wanted to investigate the effects of learning during the course of the study between protocols. To do this, we averaged user errors in the practice section and compared them with the corresponding errors in testing.

Table~\ref{tab:error_protocol} shows the average error per selection for each protocol in the two sections of the study. Notably, the \textit{Actionable} group exhibited the smallest error and the largest improvement between sections, while the \textit{Placebic} group had the largest error and the smallest improvement.

\begin{table}
    \centering
    \caption{Average error per selection across sections.}
    \label{tab:error_protocol}
    \begin{tabular}{|l|r|r|r|}
    \hline
        Protocol & Practice & Testing & \% Diff. \\
    \hline
        None & 3.23 & 2.87 & -11.3 \\
        Placebic & 3.36 & 3.05 & -9.3 \\
        Actionable & 2.95 & 2.52 & -14.5 \\
    \hline
    \end{tabular}
\end{table}

\subsection{Survey Responses}

\begin{figure}
    \centering
    \begin{subfigure}{.49\textwidth}
        \centering
        \includegraphics[width=\textwidth]{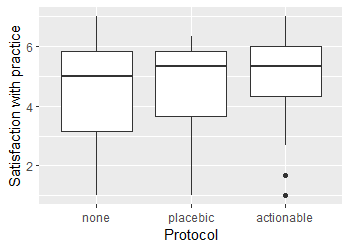}
        \caption{Satisfaction with practice.}
        \label{fig:boxplot_sat_practice_vs_error}
    \end{subfigure}
    \hspace{10pt}
    \begin{subfigure}{.49\textwidth}
        \centering
        \includegraphics[width=\textwidth]{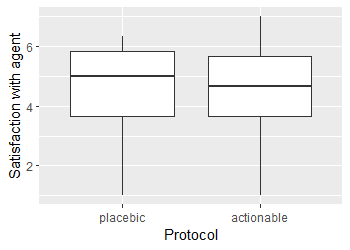}
        \caption{Satisfaction with agent.}
        \label{fig:boxplot_sat_agent_vs_error}
    \end{subfigure}
    \begin{subfigure}{.49\textwidth}
        \centering
        \includegraphics[width=\textwidth]{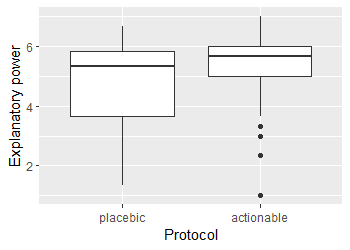}
        \caption{Explanatory power.}
        \label{fig:boxplot_exp_power_vs_error}
    \end{subfigure}
    \caption{Distributions of the three subjective survey metrics across protocols.}
    \label{fig:boxplot_survey_vars_vs_protocol}
\end{figure}

\begin{figure}
    \centering
    \includegraphics[width=.49\textwidth]{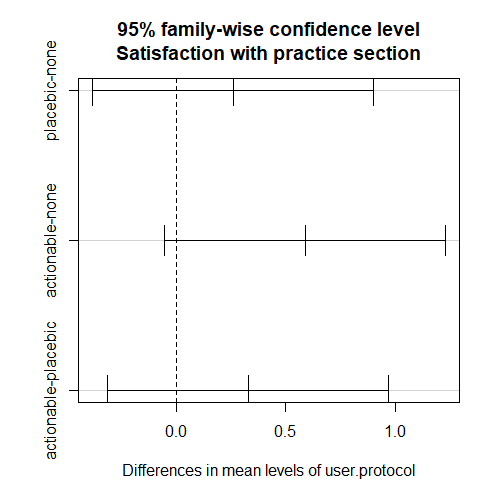}
    \caption{Tukey's HSD results for satisfaction with practice.}
    \label{fig:tukey_sat_practice}
\end{figure}

The distributions of the three subjective metrics that we collected can be found in Figure~\ref{fig:boxplot_survey_vars_vs_protocol}. We ran ANOVA to test for differences in the responses of the survey between conditions. The first was the average user satisfaction with the practice section, which exhibited an insignificant difference between the groups ($F(2, 186)=2.342$, $p=.099$, $\eta^2=.02$). The results of Tukey's HSD, seen in Figure~\ref{fig:tukey_sat_practice}, demonstrated that the \textit{Actionable} protocol did not receive significantly higher satisfaction scores than the \textit{None} protocol ($p=.0811$). \textit{Placebic} vs. \textit{None} ($p=.6074$) and \textit{Actionable} vs. \textit{Placebic} ($p=.4511$) demonstrated insignificant differences.

Next, we performed an ANOVA on user satisfaction with the agent and explanatory power. The former comparison did not reveal significance ($F(1, 124)=0.04$, $p=.843$, $\eta^2<.001$).  The latter comparison also did not show a significant difference ($F(1, 124)=2.894$, $p=.0914$, $\eta^2=.02$). It is important to note that these tests did not include the \textit{None} group since questions about the explanations would not make sense to users who did not receive explanations.

Again, we utilized TOST to indicate if there were any pairwise equivalences in the survey responses between protocols. For all subjective metrics, we set the EqB to 0.5, since all responses existed on a 7-point Likert scale. We found that the \textit{Placebic} and \textit{Actionable} participants were not equally satisfied with their practice experiences ($t(124)=0.668$, $p=.253$, Hedges' $g=-.23$). However, they were equally satisfied with their respective agents ($t(124)=1.681$, $p=.048$, Hedges' $g=-.04$). There was no equivalence found between their agents' perceived explanatory power ($t(124)=0.213$, $p=.416$, Hedges' $g=-.30$).

We summarize all of our one-way ANOVA test results in Table~\ref{tab:anova_results}, and all of the TOST results can be found in Table~\ref{tab:tost_results}.

\begin{table}
    \centering
    \caption{ANOVA results comparing user performance and survey responses across explanation protocols. Bold $p$-values show significant differences.}
    \label{tab:anova_results}
    \begin{tabular}{|l|r|r|r|r|}
    \hline
        Metric & $df$ & $F$ & $p$ & $\eta^2$ \\
    \hline
        Testing error & (2, 186) & 5.89 & \textbf{.003} & .06 \\
        Bonus compensation & (2, 186) & 5.26 & \textbf{.006} & .05 \\
    \hline
        Satisfaction with practice & (2, 186) & 2.34 & .099 & .02 \\
        Satisfaction with agent & (1, 124) & 0.04 & .843 & <.01 \\
        Explanatory power & (1, 124) & 2.89 & .091 & .02 \\
    \hline
    \end{tabular}
\end{table}

\begin{table}
    \centering
    \caption{TOST results comparing user performance and survey responses across explanation protocols. Bold $p$-values show significant equivalences.}
    \label{tab:tost_results}
    \begin{tabular}{|l|l|r|r|r|r|r|}
    \hline
        Metric & Comparison & EqB & $df$ & $t$ & $p$ & Hedges' $g$ \\
    \hline
        Testing error & \textit{None-Placebic} & .5 & 124 & 1.821 & \textbf{.036} & -.23 \\
        Bonus compensation & \textit{None-Placebic} & .1 & 124 & -3.137 & \textbf{.001} & .11 \\
    \hline
        Satisfaction with practice & \textit{Placebic-Actionable} & .5 & 124 & 0.668 & .253 & -.23 \\
        Satisfaction with agent & \textit{Placebic-Actionable} & .5 & 124 & 1.681 & \textbf{.048} & -.04 \\
        Explanatory power & \textit{Placebic-Actionable} & .5 & 124 & 0.213 & .416 & -.30 \\
    \hline
    \end{tabular}
\end{table}

\subsection{Correlation}

\begin{figure}
    \centering
    \begin{subfigure}{.49\textwidth}
        \centering
        \includegraphics[width=\textwidth]{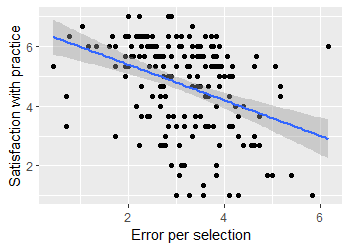}
        \caption{Satisfaction with practice.}
        \label{fig:cor_sat_practice_vs_error}
    \end{subfigure}
    \hspace{20pt}
    \begin{subfigure}{.49\textwidth}
        \centering
        \includegraphics[width=\textwidth]{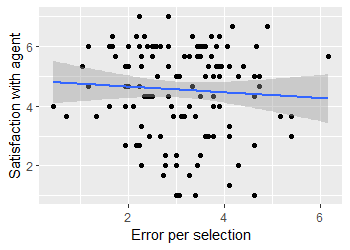}
        \caption{Satisfaction with agent.}
        \label{fig:cor_sat_agent_vs_error}
    \end{subfigure}
    \begin{subfigure}{.49\textwidth}
        \centering
        \includegraphics[width=\textwidth]{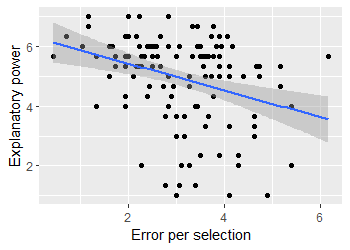}
        \caption{Explanatory power.}
        \label{fig:cor_exp_power_vs_error}
    \end{subfigure}
    \caption{Survey metrics vs. error.}
    \label{fig:cor_survey_vars_vs_error}
\end{figure}

Finally, we were interested in extracting information about the relationship between our surveyed subjective metrics (satisfaction with practice, satisfaction with agent, and perceived explanatory power of explanations) and user performance. Figure~\ref{fig:cor_survey_vars_vs_error} plots the three metrics against each user's mean testing error per selection. A linear trend line was fitted to each plot along with a 95\% confidence band.

\begin{table}
    \centering
    \caption{Correlation between error and survey metrics. Bold $p$-values show significant correlations.}
    \label{tab:cor_test}
    \begin{tabular}{|l|r|r|r|r|}
    \hline
        Metric & $df$ & $t$ & $p$ & $r$ \\
    \hline
        Satisfaction with practice & 187 & -5.813 & \textbf{<.001} & -0.391 \\
        Satisfaction with agent & 124 & -0.755 & .452 & -0.068 \\
        Explanatory power & 124 & -3.772 & \textbf{<.001} & -0.321 \\
    \hline
    \end{tabular}
\end{table}

We ran three tests for correlation between samples using the Pearson correlation coefficient as the statistic. These results are summarized in Table~\ref{tab:cor_test}. The correlations involving satisfaction with the practice section and explanatory power were statistically significant ($p<.001$), while satisfaction with the agent was not significantly correlated with error ($p=.452$).

\section{Discussion}

Hypothesis 1 (H1) predicted that actionable explanations would lead to significantly better user performance than placebic explanations. The data strongly support this hypothesis, as shown by Tukey's HSD results for both error rate ($p=.0025$) and bonus compensation ($p=.0075$). Participants in the \textit{Actionable} group were more accurate and earned higher bonuses, indicating that actionable explanations offer real value in terms of performance enhancement. Moreover, the \textit{Actionable} group demonstrated the largest improvement between the practice and testing sections (Table~\ref{tab:error_protocol}), which emphasizes the inherent ability of actionable explanations to promote task understanding and execution. This result reinforces the argument that explanations that provide actionable insights are more effective in enhancing model understandability and user decision-making.

H2 proposed that placebic explanations would not lead to better user performance compared to the absence of explanations. This was confirmed by the data, as there was a significant equivalence between the \textit{Placebic} and \textit{None} groups' error rate ($p=.036$) and the bonus compensation ($p=.001$). The lack of improvement in the \textit{Placebic} group suggests that placebic explanations are essentially ineffective, offering no advantage over receiving no explanation at all. This finding places placebic explanations as a low-quality counterpart to actionable explanations.

H3 asserted that actionable explanations would not significantly increase user satisfaction compared to placebic explanations. We showed that actionable and placebic explanations provide equivalent user satisfaction with the agent ($p=.048$). While the same equivalence could not be demonstrated for satisfaction with the practice section ($p=.253$), performance was found to be significantly correlated with this metric. Thus, we believe that the data support this hypothesis. This result highlights the limitations of subjective satisfaction as a metric for evaluating the quality of the explanation. Users are just as satisfied with placebic explanations as they are with actionable ones, despite the substantial differences in performance outcomes in these two situations. This underscores the risk of relying solely on satisfaction surveys to assess the effectiveness of XAI systems.

H4 suggested that placebic explanations would lead to higher user satisfaction than without explanations. However, the data do not support this hypothesis, as there was no significant difference in satisfaction between the \textit{Placebic} and \textit{None} groups ($p=.6074$). This finding is particularly interesting, as it seems to challenge the assumption that any explanation, even a low-quality one, will satisfy users more than receiving no explanation would. Perhaps there were too many other satisfying design elements in the study for the introduction of explanations to cause a significant increase.

H5 stated that actionable explanations would not lead to significantly higher perceived explanatory power compared to placebic explanations. The data do not support this hypothesis. Neither a difference ($p=.091$) nor an equivalence ($p=.416$) was found to be significant between the groups. Thus, there is a nuanced relationship between these types of explanations and their perceived power by the user, requiring more results to make a definitive statistical statement.

Comparing the correlation results in Table~\ref{tab:cor_test} with the ANOVA results in Table~\ref{tab:anova_results} reveals that the subjective responses of the users were more closely related to their performance than to the explanation protocol. Regardless of the quality of the explanation, the users felt more satisfied and perceived the explanations to be more powerful when they performed well. This suggests that user satisfaction and perceived explanatory power are not reliable indicators of explanation quality—they are influenced by how well users perform, independent of the explanations they receive. Interestingly, user satisfaction with the agent was not affected by either performance or protocol, raising concerns that low-quality explanations can satisfy users as much as high-quality ones, as long as they are presented clearly and seem relevant. This highlights a significant issue for the design of the XAI system: Users may be content with explanations that do not improve their understanding or performance.

\section{Conclusions}

These observations suggest a disconnect between subjective measures such as satisfaction and objective performance measures in evaluating XAI systems. Although users may relate satisfaction with superficial attributes of the agent or experience (regardless of performance), actionable explanations result in measurable objective benefits. This finding reinforces the idea that subjective user satisfaction alone is insufficient to assess the quality of the explanation and should be supplemented with objective metrics.

We found that satisfaction only weakly correlated with high performance on tests. 
This demonstrates that satisfaction is not a suitable replacement for demonstrating concept understanding. Performance measurement through testing was more effective at identifying the actionable explanations, while users reported satisfaction with placebic explanations.

We believe these results are particularly relevant to agent experts, such as pedagogical agents. In the Social Security optimization scenario, users lacked experience understanding the domain. In this scenario, the agent is an expert, explaining new concepts to users. Our findings may apply most strongly to pedagogical agents. 

\section{Limitations and Future Work}

Combined with previous research results on other subjective metrics, the results of this study support the strong claim that subjective metrics based on user responses should not be relied upon in the absence of objective metrics. However, there are some limitations that need to be addressed in future work.

We focus on reported satisfaction as a metric. Although it is clear that there is a strong correlation among other subjective metrics such as trust, confidence, reported understandability, and others, there may be cases where users respond differently to these metrics.  

The results of this study are limited to users who have neither machine learning expertise nor domain knowledge.  There is evidence that subjective metrics are more powerful when provided by experts in the task domain~\cite{ford2022explaining,lopes2022xai}. More research on the relationship between domain expertise and evaluation techniques is needed to confirm this hypothesis.

The goal of the explanations in this study is to increase the user's understanding about an automated decision process in an unfamiliar environment. Although understandability is the most common goal of explanations, there have been an increasing number of alternative primary goals for explanation systems~\cite{mohseni2021multidisciplinary}. In particular, the number of studies that involve human-AI teams with the goal of optimizing team performance on a task has increased substantially over the past few years~\cite{lopes2022xai}. Future works may consider the role that alternative explanation goals have when choosing evaluation metrics. 

There may be significant differences between different types of placebic explanations. We primarily used restatements of known instructional facts, but alternative placebic explanations could include tautologies or off-topic statements. We consider random, misleading, or outright false explanations to be completely separate and generally not appropriate as a baseline. However, comparing user responses for different kinds of "bad" explanations with placebic explanations and good actionable explanations would be valuable for deciding appropriate evaluation metrics.

\appendix

\section{Example Explanations}
\label{sec:example_explanations}

Table~\ref{tab:example_explanations} provides examples of actionable and placebic explanations in a few different categories of hypothetical Social Security scenarios.

\begin{table}
    \centering
    \caption{Several examples of each explanation type for identical scenarios. The scenario codes describe the optimal filing ages, single (s) or married (m) individuals, and whether spousal benefits (ms) affect the decision.}
    \label{tab:example_explanations}
    \begin{tabularx}{\textwidth}{|l|X|X|}
    \hline
        Scenario & Placebic & Actionable \\
    \hline
        70s & Filing at 70 produces the most benefits in the long run for this individual. & This person has a long lifespan, so they should file later to increase their monthly payments. \\
        62m & The optimal filing ages ensure that this couple will get the most from Social Security over their lifetimes. Each spouse's PIA, age, and lifespan work together to determine the best times to file. & Both spouses have short lifespans, so they should file earlier to increase the number of years they receive payments. \\
        67-70ms & The ages that these people should file at depends on the factors provided: PIA, age, expected longevity and marital status. If Spouse A files for benefits at 67 years old while Spouse B files when they turn 70, the couple as a whole will actually receive the best payout possible from Social Security. & Both spouses have long lifespans, so they should file later to increase their monthly payments. The two PIAs are significantly different, so the lower PIA spouse (Spouse A) will receive spousal benefits. Spousal benefits do not increase past 67, so Spouse A should file then. \\
    \hline
    \end{tabularx}
\end{table}

\section{Survey Statements}
\label{sec:survey_statements}

Table~\ref{tab:survey_statements} shows the exact statements used in our final surveys. Users responded to each with their level of agreement on a 7-point Likert scale. The results were then averaged to obtain a single representative score for each metric.

\begin{table}
    \centering
    \caption{Final survey statements given to participants upon completion of the testing section. The respondents' agreement levels came from a 7-point Likert scale. Those in the \textit{None} protocol did not respond to the latter two categories, as they did not receive explanations from an AI agent.}
    \label{tab:survey_statements}
    \begin{tabularx}{\textwidth}{|l|X|}
    \hline
        Category & Statement \\
    \hline
        Satisfaction with practice & The practice section prepared me well for the test section. \\
        Satisfaction with practice & I was well equipped to tackle the test section after completing the practice section. \\
        Satisfaction with practice & Without the practice section, I would have performed worse in the test section. \\
        \hline
        Satisfaction with agent & I enjoyed the AI agent. \\
        Satisfaction with agent & I was satisfied with the AI agent's explanations. \\
        Satisfaction with agent & I found the collaboration with the AI agent to be enjoyable. \\
        \hline
        Explanatory power & The agent's explanations helped me understand the scenarios better. \\
        Explanatory power & Without the agent's explanations, the scenarios would have been tougher. \\
        Explanatory power & The explanations from the agent gave me helpful insight to solve future scenarios. \\
    \hline
    \end{tabularx}
\end{table}

\begin{credits}
\subsubsection{\ackname} The authors were supported by the Cyber Fellows program at The University of Tulsa.

\subsubsection{\discintname} The authors have no competing interests to declare that are relevant to the content of this article.
\end{credits}

\bibliographystyle{splncs04}
\bibliography{socsec/references}

\end{document}